\documentclass{zHenriquesLab-StyleBioRxiv}
\usepackage{blindtext}
\usepackage{lipsum}
\leadauthor{Kelam} 

\begin{document}

\title{Ensemble YOLO Framework for Multi-Domain Mitotic Figure Detection in Histopathology Images}
\shorttitle{Approach for MIDOG 2025}
\author{Navya Sri Kelam}
\author{Akash Parekh}
\author{Saikiran Bonthu} 
\author{Nitin Singhal}

\affil{AIRA MATRIX Private Limited, Mumbai, India}
\maketitle
\begin{abstract}
The reliable identification of mitotic figures in whole-slide histopathological images remains difficult, owing to their low prevalence, substantial morphological heterogeneity, and the inconsistencies introduced by tissue processing and staining procedures. The MIDOG competition series provides standardized benchmarks for evaluating detection approaches across diverse domains, thus motivating the development of generalizable deep learning models. In this work, we investigate the performance of two modern one-stage detectors, YOLOv5 and YOLOv8, trained on MIDOG++, CMC, and CCMCT datasets. To enhance robustness, training incorporated stain-invariant color perturbations and texture-preserving augmentations. In internal validation, YOLOv5 achieved higher precision (84.3\%), while YOLOv8 offered improved recall (82.6\%), reflecting architectural trade-offs between anchor-based and anchor-free detections. To capitalize on their complementary strengths, we employed an ensemble of the two models, which improved sensitivity (85.3\%) while maintaining competitive precision, yielding the best F1 score of 83.1\%. On the preliminary MIDOG 2025 test leaderboard, our ensemble ranked 5th with an F1 score of 79.2\%, precision of 73.6\%, and recall of 85.8\%, confirming that the proposed strategy generalizes effectively across unseen test data. These findings highlight the effectiveness of combining anchor-based and anchor-free object detectors to advance automated mitosis detection in digital pathology.
\end{abstract}


\begin{corrauthor}
nitin.singhal@airamatrix.com
\end{corrauthor}

\section*{Introduction}
Accurate mitosis detection is essential for histopathological diagnosis, directly impacting tumor classification and treatment planning. Challenges arise from the rarity of mitotic figures, their resemblance to non-mitotic structures, and staining variability across laboratories. The Medical Image Computing and Computer-Assisted Intervention (MICCAI) Mitosis Domain Generalization Challenge (MIDOG) \cite{midog2022} aimed to foster algorithms that generalize accurately across domains. Automated mitosis detection has advanced rapidly, largely due to the MIDOG challenges, which benchmark cross-domain performance. MIDOG 2021 \cite{midog2021} focused on breast cancer, enabling domain-specific generalization but limiting robustness. MIDOG++\cite{aubreville_comprehensive_2023} and MIDOG 2025 \cite{midog2025} introduced multi-domain datasets with diverse tissues, scanners, and staining protocols, making generalization a key challenge. Deep learning-based object detectors, especially the YOLO family, are widely used for mitosis detection due to their efficiency and accuracy. This study compares YOLOv5 and YOLOv8 on a curated multi-center dataset, examining how augmentation, dataset design, and architecture affect precision–recall dynamics, with a focus on model generalizability in heterogeneous tissue settings.

\section*{Related Work}
The MIDOG challenge series has been pivotal in benchmarking mitosis detection algorithms under domain shift conditions. The MIDOG 2021 challenge \cite{midog2021} focused exclusively on breast cancer tissue and highlighted that domain generalization arising from variability in tissue preparation, laboratory protocols, and scanner technologies remains a central challenge in histopathological mitosis detection. While participants achieved high F1 scores, performance deteriorated on images from unseen scanners or conditions. Building on this, the MIDOG 2022 challenge \cite{midog2022} expanded to include multiple tumor types, allowing the evaluation of generalization strategies across more diverse histopathological domains. Participants were allowed to use publicly available mitosis datasets in addition to official training data, but the report noted that domain shifts continued to challenge algorithmic performance. Further extending these efforts, the MIDOG++ dataset and benchmark \cite{aubreville_comprehensive_2023} incorporated a wider variety of tumor types and scanner modalities, and investigated training strategies such as leave-one-domain-out experiments to systematically assess domain generalization. These studies collectively underscore that increasing the diversity of training data is crucial for achieving robust and transferable mitosis detection performance across heterogeneous histopathological settings.

\section*{Datasets and Preprocessing}
This study leverages three mitosis detection datasets to evaluate domain generalization and cross-species robustness. The MIDOG25 dataset\cite{midog2025} spans seven domains, including human and canine tumors, providing a diverse multi-scanner, multi-tumor benchmark for generalization. The \textbf{MITOS\_WSI\_CMC} dataset \cite{aubreville_completely_cmc_2020}, focused on canine mammary carcinoma, introduces a cross-species domain shift. Meanwhile, the \textbf{MITOS\_WSI\_CCMCT} dataset \cite{aubreville_completely_ccmct_2020}, comprising canine cutaneous mast cell tumor slides, adds further complexity through interspecies and tumor-type variability. Following the MIDOG25 challenge protocol, the MIDOG25 dataset was first divided into training and testing subsets, with 80\% of ROIs allocated for training and 20\% reserved for evaluation. For the CMC and CCMCT datasets, three to five whole-slide images per center were selectively chosen to ensure histological heterogeneity, including tumor-dense regions, fibrotic tissue, necrotic zones, and normal stroma.

\section*{Augmentation}
Augmentation was pivotal in this study, given the sensitivity of deep learning models in digital pathology to staining and imaging variability. Drawing on Litjens et al. (2017) \cite{tellez2018he_stain_augmentation}, we implemented a diverse set of augmentations that simulate realistic conditions while maintaining the morphological integrity of the mitotic figures. To address staining heterogeneity across labs, we applied color augmentations, specifically hue, saturation, and brightness shifts, to mimic variations in hematoxylin and eosin concentrations. These transformations help models prioritize structural features over raw color intensity. Texture-based augmentations were introduced to reflect acquisition-level inconsistencies. Gaussian blur simulated out-of-focus regions caused by tissue thickness or scanner optics, while sharpening enhanced nuclear edges to aid in distinguishing mitotic chromatin from apoptotic or necrotic nuclei. Gaussian noise was added to replicate scanner noise and preparation artifacts. We also incorporated advanced compositional augmentations. Mosaic augmentation \cite{bochkovskiy2020yolov4} combined four random images into a single composite, enabling the model to learn across varied scales and contexts, especially useful for detecting rare events like mitoses. Cutmix \cite{yun2019cutmix} exposed the model to partially visible nuclei, improving robustness in cases where mitotic figures are only partially captured. Together, these augmentations enhanced the model’s ability to generalize across domains, making it more resilient to the diverse challenges encountered in real-world histopathology.

\section*{Model Architectures}
\textbf{YOLOv5-l}\cite{khanam2024what_is_yolov5} employs a CSPDarknet\cite{wang2020cspnet} backbone with an anchor-based detection head. As an anchor-based model, its predictions are tightly coupled to predefined anchor priors, often resulting in conservative detection behavior with higher precision.

\textbf{YOLOv8-m}\cite{ultralytics_yolov8} introduces several critical improvements over its predecessor. Most notably, it adopts an anchor-free detection paradigm with a decoupled head, separating classification and localization tasks. This design not only reduces computational overhead but also improves the ability of the model to recall rare and small objects. Enhanced feature aggregation within YOLOv8-m further facilitates gradient propagation, which is particularly beneficial for fine-grained mitotic detection. Collectively, these architectural refinements enable YOLOv8-m to achieve higher recall, although at the cost of slightly reduced precision compared to YOLOv5-l.

\section*{Model Training and Ensembling}
Both YOLOv5-l and YOLOv8-m were trained independently for 200 epochs with an input resolution of 1024 × 1024. The Adam optimizer was employed with an initial learning rate of 0.01, decayed using a cosine annealing scheduler to stabilize convergence. Loss functions included bounding box regression, classification loss, and distribution focal loss. To further improve training stability and model performance, hyperparameters such as learning rate, augmentation probabilities, and loss coefficients were optimized using the Ultralytics tune module \cite{ultralytics_tune}. This automated tuning procedure enabled a systematic exploration of the hyperparameter space, ensuring that both models were trained under near-optimal conditions.

The choice of YOLOv5 and YOLOv8 was motivated by their architectural differences, which yield complementary detection behaviors. YOLOv5 which employs a CSPDarknet backbone with a PANet neck and an anchor-based detection head, optimized with CIoU and binary cross-entropy losses. YOLOv8, in contrast, introduces a CSP-Next backbone\cite{chen2024cspnext} with a decoupled head, adopts an anchor-free detection paradigm, and leverages advanced objectives such as the task aligned assigner and the distribution focal loss. These differences result in distinct inductive biases: YOLOv5 tends to favor well-structured mitotic figures that conform to anchor priors, producing more conservative predictions with higher precision, whereas YOLOv8 exhibits greater flexibility in localizing smaller or irregular mitoses, often leading to higher recall.

\begin{figure*}[tbhp]
\centering
\includegraphics[width=\linewidth]{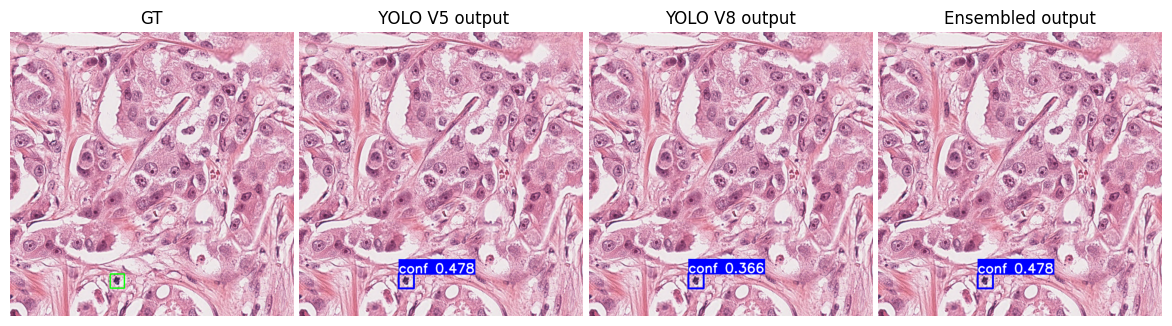}
\caption{Example illustrating tightly fitting mitotic figures. YOLOv5 successfully detected the instance with a high-confidence bounding box (left), whereas YOLOv8 failed to identify it (middle). The ensembled output (right) retained the YOLOv5 prediction after non-maximum suppression, highlighting the complementarity between the two models.}
\label{fig:tightfit}
\end{figure*}

\begin{figure*}[tbhp]
\centering
\includegraphics[width=\linewidth]{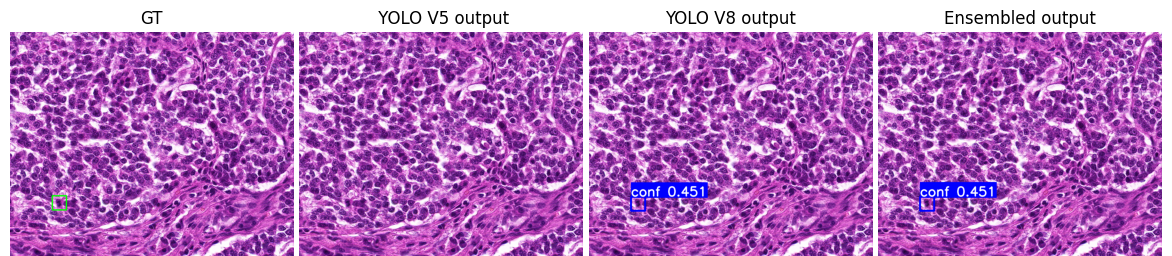}
\caption{Example illustrating fine-grained mitotic figures. YOLOv5 failed to predict the instance (left), while YOLOv8 successfully detected it (middle). The ensembled output (right) retained the YOLOv8 prediction after non-maximum suppression, demonstrating the benefit of combining the two models.}
\label{fig:finegrained}
\end{figure*}

Model performance was assessed using precision, recall, and the F1 score, as these jointly capture the trade-off between false positives and false negatives in the highly imbalanced mitosis detection task. The results showed that YOLOv5 provided greater precision, while YOLOv8 offered improved recall, confirming their complementary strengths. Specifically, YOLOv5 excelled at detecting familiar objects that fit tightly within bounding boxes (Figure~\ref{fig:tightfit}), often assigning them high confidence scores, but it struggled with small and rare objects. In contrast, YOLOv8 demonstrated superior performance on small and rare objects (Figure~\ref{fig:finegrained}), making it a strong complement to YOLOv5. To exploit this complementarity, the final submission employed an ensemble of YOLOv5 and YOLOv8, combining their predictions to achieve higher sensitivity without substantially sacrificing precision. First, predictions from both models were obtained independently for each image patch using a decision threshold of 0.399. The predicted bounding boxes were then mapped back to global slide coordinates by adding tile offsets. All predictions from the two models were subsequently aggregated into a single candidate set. To eliminate duplicate detections of the same object, non-maximum suppression (NMS) with an IoU threshold of 0.4 was applied to the combined predictions, and finally, the box with the highest confidence score was retained.

\section*{Results and Discussion}
In our experiments, YOLOv5-l achieved a precision of 0.8431 and a recall of 0.7928, reflecting its more conservative detection strategy. This behavior can be attributed to its anchor-based design: Since the ground-truth boxes are of uniform size, the anchor priors converge tightly to this scale, resulting in higher confidence for detections that align well with the anchor while occasionally missing less well-conforming instances. In contrast, YOLOv8-m, being anchor-free, demonstrated greater flexibility in localizing objects across subtle intra-box size variations, yielding a higher recall of 0.8261. The complementary nature of these detection paradigms motivated our ensemble strategy, which improved overall sensitivity while maintaining competitive precision. On the preliminary MIDOG 2025 leaderboard, our ensembled method was ranked 5th with an F1 score of 0.7923, precision of 0.7357, and recall of 0.8583 (Table~\ref{tab:results}). These results are consistent with internal validation, underscoring that ensembling balances YOLOv5’s precision bias with YOLOv8’s recall-oriented design. Compared to higher-ranked entries, our method demonstrates competitive recall, confirming the robustness of anchor–anchor-free ensembling under multi-domain conditions.

\begin{table}[ht]
\centering
\begin{tabular}{lccccc}
\hline
\textbf{Model / Submission} & \textbf{Precision} & \textbf{Recall} & \textbf{F1 Score}\\
\hline
YOLOv5-l (internal)   & 84.31 & 79.28 & 81.71\\
YOLOv8-m (internal)   & 82.87 & 82.61 & 82.74\\
Ensemble (internal)   & 81.11 & \textbf{85.25} & \textbf{83.13}\\
Ensemble (Leaderboard) & \textbf{73.57} & \textbf{85.83} & \textbf{79.23}\\
\hline
\end{tabular}
\caption{Performance of YOLOv5-l, YOLOv8-m, and their ensemble on internal validation, and the ensemble submission on the preliminary MIDOG 2025 leaderboard.}
\label{tab:results}
\end{table}

\section*{Acknowledgements}
The authors acknowledge the organizers of the MIDOG25 challenge for providing a well-structured platform to advance research in medical imaging diagnostics. Their efforts in acquiring high-quality datasets and fostering collaborative innovation have been instrumental to this work.

We also extend our sincere thanks to my organization Airamatrix pvt ltd for supporting and encouraging our participation in this challenge. The opportunity to contribute to this initiative reflects the organization's commitment to driving impactful solutions in AI-powered healthcare.

\section*{References}
\sloppy 
\bibliography{literature}

\end{document}